\documentclass[conference]{IEEEtran}
\usepackage{graphicx}
\usepackage{amsmath}
\usepackage{cite}
\usepackage{listings}
\usepackage{xcolor}
\usepackage{url}

\begin{document}

\title{Phishing URL Detection using Bi-LSTM}

\author{
    \IEEEauthorblockN{Sneha Baskota}
    \IEEEauthorblockA{
        The University of Texas Permian Basin
    }
}

\maketitle

\begin{abstract}
Phishing attacks continue to be a significant threat to online users, leading to data breaches, financial losses, and identity theft. Traditional phishing detection systems struggle with high false positive rates and are often limited by the types of attacks they can identify. This paper proposes a deep learning-based approach using a Bidirectional Long Short-Term Memory (Bi-LSTM) network to classify URLs into four categories: benign, phishing, defacement, and malware. The model leverages sequential URL data and captures contextual information, improving the accuracy and robustness of phishing detection. Experimental results on dataset comprising over 650,000 URLs demonstrate the effectiveness of the model, achieving 97\% accuracy and significant improvements over traditional techniques.
\end{abstract}

\IEEEpeerreviewmaketitle

\section{Introduction}
Phishing attacks, which deceive users into providing sensitive information by impersonating legitimate entities, continue to be one of the most prevalent forms of cybercrime. According to recent reports, phishing attacks account for a substantial portion of cybersecurity incidents worldwide, with millions of users falling victim annually \cite{ASIRI2024103843}. Traditional phishing detection methods, including rule-based and heuristic approaches, are increasingly ineffective as attackers employ more sophisticated techniques. This paper aims to enhance phishing URL detection by introducing a deep learning model that leverages a Bi-LSTM architecture. This approach targets not only phishing URLs but also includes defacement and malware URLs, expanding the scope of detection and providing a more robust solution for real-world applications.
\begin{figure}[h!]
    \centering
    \includegraphics[width=0.8\linewidth]{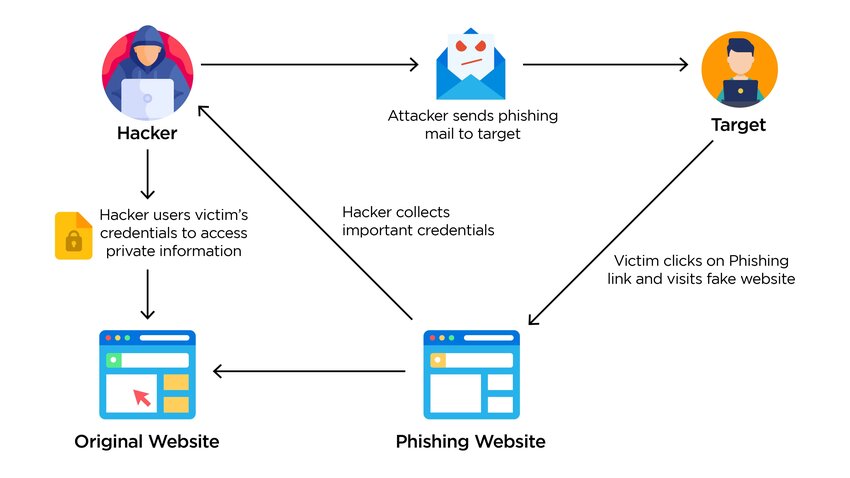}  
    \caption{Phishing attack process}
    \label{fig:phishing attack process}
\end{figure}

\section{Dataset and Preprocessing}
The dataset used for this project is sourced from Kaggle, which contains a diverse set of URLs classified into four categories: benign, phishing, defacement, and malware. The benign URLs represent safe and legitimate web addresses, while phishing URLs are those designed to deceive users and steal sensitive data. Defacement URLs correspond to sites that have been altered by attackers, often displaying malicious content. Malware URLs host harmful software, intended to infect user systems.

The data set includes over 650,000 URLs, with a distribution of 428,103 benign URLs, 94,110 phishing URLs, 96,456 defacement URLs, and 32,520 malware URLs. Preprocessing steps involved tokenizing the URLs into characters and padding them to a fixed length, ensuring compatibility with the Bi-LSTM model. In addition, URLs were encoded into numerical representations and categorical labels were assigned to each URL based on its classification.

\section{Literature Review}
Phishing detection has been an area of significant research in cybersecurity, with various approaches being proposed over the years. The early methods relied on manual feature extraction and rule-based classifiers. However, these techniques struggled to handle large-scale data sets and the evolving nature of phishing tactics. Recent advancements have incorporated machine learning algorithms, such as decision trees, support vector machines (SVMs), and random forests, which improve detection rates by learning patterns from historical data \cite{ghalechyan2024phishing}. 

Deep learning models, particularly recurrent neural networks (RNNs) and convolutional neural networks (CNNs), have recently gained popularity for phishing detection tasks due to their ability to capture complex patterns in sequential data\cite{article}. Bi-LSTM networks, a variant of RNNs, have shown great promise in sequential data analysis, as they capture past and future context in a given sequence\cite{zhou2025integrated}. Furthermore, attention mechanisms have been integrated into Bi-LSTM models to enhance their ability to focus on important features in input data \cite{phishing_bilstm_attention}.

Several studies have applied these techniques to phishing detection with varying levels of success. For example, Newaz \textit{et al.} proposed a combination of feature selection, greedy algorithm, cross-validation, and deep learning methods to construct a sophisticated stacking ensemble classifier\cite{newaz2024sophisticatedframeworkaccuratedetection}. However, their approach focused primarily on binary classification (benign vs. phishing). In contrast, my work extends this by incorporating multiclass classification, including defacement and malware categories, providing a more comprehensive solution. Nanda et al. proposed a model that integrates the highway network into the BiLSTM-CNN architecture, which enables the capture of significant features with rapid convergence\cite{nanda2024url}. 

Fajar \textit{et al.} explored phishing URL detection using machine learning models, emphasizing the impact of feature selection and model interpretability~\cite{fajar2024enhancingphishingdetectionfeature}. Their study employed recursive feature elimination to identify key indicators, including \texttt{length\_url}, \texttt{time\_domain\_activation}, and \texttt{Page\_rank}, which significantly enhanced classification performance. Among the evaluated models, XGBoost demonstrated high efficiency with respect to runtime, making it suitable for large-scale datasets, while CatBoost maintained strong accuracy even with a reduced feature set. To improve transparency, the study incorporated explainable AI techniques such as SHAP, providing interpretable insights into feature importance and highlighting the value of integrating explainability into phishing detection systems.

Also, the 1D Convolutional Neural Network (CNN) model proposed by Islam et al. uses a very large dataset and extracts 21 features with a 99\% accuracy\cite{islam2024phishguardconvolutionalneuralnetwork}.

The authors in Asiri \textit{et al.} proposed a system that detects three types of phishing attacks:Tiny Uniform Resource Locators (TinyURLs), Browsers in the Browser (BiTB), and regular phishing attacks \cite{ASIRI2024103843}. They used a deep learning model along with a browser extension that would prevent users from opening malicious links. The models proposed by Jishnu \textit{et al.} were complex and time-consuming\cite{jishnu2025phishing}.

\section{Model Architecture}

The core of our phishing URL detection system is built upon a \textit{Bidirectional Long Short-Term Memory (Bi-LSTM)} neural network. This architecture is well-suited for sequence modeling tasks, especially those involving textual or character-level data such as URLs.

\subsection{Input Layer}
The input to our model consists of tokenized URLs. Each URL is treated as a sequence of individual characters. These characters are first tokenized and then converted into integer indices, allowing the model to process them numerically. To ensure uniformity across all input samples, each sequence is padded to a fixed maximum length.

\subsection{Embedding Layer}
The first learnable layer of the network is an embedding layer, which transforms each character index into a dense vector of fixed size. This allows the model to learn distributed representations of characters, capturing semantic similarities and contextual patterns. For example, characters like \texttt{'/'}, \texttt{'?'}, and \texttt{'.'}, which often occur in specific URL structures, may be embedded in similar regions of the vector space.

\subsection{Bi-LSTM Layer}
Following the embedding layer, a Bidirectional LSTM (Bi-LSTM) processes the sequence. Unlike a traditional LSTM that reads the sequence in a single direction, a Bi-LSTM reads the input both forward and backward, and then concatenates the two outputs. This allows the model to learn contextual information from both past and future characters, which is crucial to detect complex phishing patterns.

The LSTM component itself is capable of learning long-term dependencies in sequential data, helping the model understand patterns such as suspicious domain names, common phishing prefixes or suffixes, and malformed URL structures.

\subsection{Dropout Layer}
To prevent overfitting and enhance generalization, a dropout layer is applied after the Bi-LSTM layer. This randomly drops a fraction of the connections during training, ensuring that the model does not rely too heavily on any particular pathway and memorizes the patterns instead of learning the patterns. 

\subsection{Dense Layer and Output}
The final stage includes one or more fully connected (dense) layers. The last dense layer uses a \textit{softmax} activation function to output a probability distribution over the predefined URL categories: \textit{benign, phishing, defacement}, and \textit{malware}. The model is trained to minimize \textit{categorical cross-entropy loss}, which encourages the predicted probability to be high for the correct class.

\section{Flask Backend and User Interface}

To make the phishing URL detection model accessible and user-friendly, I developed a web-based application using the Flask framework for the backend and a modern, interactive user interface (UI) for the front end. This section outlines the steps involved in setting up the backend with Flask and designing the UI to visualize the detection results.

\subsection{Flask Backend}

The backend of the application is powered by Flask, a lightweight Python web framework. Flask allows for the seamless integration of the trained Bi-LSTM model into a web-based application. The backend processes incoming requests from the frontend, runs the phishing URL detection model, and returns the classification results to the frontend.

\subsubsection{Model Loading and Prediction}
At the core of the Flask backend is the trained Bi-LSTM model, which is loaded using the Keras API. Upon receiving a URL to classify, the backend tokenizes and preprocesses the URL similarly to how the model was trained. The preprocessed URL is then passed through the model to obtain predictions. The results, which include the predicted class (e.g., benign, phishing, defacement, or malware) and the confidence score, are sent back to the frontend for display.

\subsubsection{Flask API Endpoint}
A RESTful API endpoint was created using Flask’s routing mechanism. This endpoint accepts POST requests containing the URLs to classify, processes the data, and returns the classification results. Below is a sample code snippet that illustrates the API endpoint for URL classification:

\begin{lstlisting}[language=Python, caption={Flask API Prediction Endpoint}]
@app.route('/predict', methods=['POST'])
def predict():
    url = request.form['url']
    # Preprocess the URL and predict
    prediction = model.predict(process_url(url))
    return jsonify({'prediction': prediction,
                    'confidence': confidence})
\end{lstlisting}

The API endpoint is designed to handle multiple URL classifications simultaneously and efficiently.

\subsection{User Interface (UI)}

The frontend of the application is built using modern web technologies, such as \textbf{HTML}, \textbf{CSS}, and \textbf{JavaScript}. The goal was to create a user-friendly UI to allow users to input URLs for classification and visualize the results.

\subsubsection{UI Features}
 Key UI elements include:
\begin{itemize}
    \item A text input field where users can enter the URL they wish to classify.
    \item An animated title with a typing effect that enhances the user experience.
    \item A button to initiate the URL classification process.
    \item A visual display of the classification result, including the predicted label (e.g., phishing or benign) and the confidence score.
    \item Background animations that add a dynamic visual appeal.
\end{itemize}

\subsubsection{Interaction Flow}
Once the user enters a URL and presses the "Scan" button, the frontend sends a POST request to the Flask backend API with the URL. The backend processes the URL, classifies it, and sends the result back. The frontend then displays the prediction result and confidence score in real time, providing users with immediate feedback.

\subsubsection{Technologies Used}
The UI was developed using:
\begin{itemize}
    \item Google Colab was used to preprocess and train the model because of its GPU support, which significantly shortened the training time.
    \item \textbf{HTML} and \textbf{CSS} for structuring and styling the webpage.
    \item \textbf{JavaScript} for implementing dynamic features and handling asynchronous communication with the backend.
    \item \textbf{Flask} for managing HTTP requests and serving the model's predictions.
    \item \textbf{AJAX} for sending data to the backend and receiving responses without requiring a full page reload.
\end{itemize}

\section{Results and Evaluation}
The model was evaluated using several performance metrics, including accuracy, precision, recall, and F1-score, to assess its ability to correctly classify URLs. Table I shows the classification report for the model's performance on the test set.

\begin{table}[h!]
\centering
\begin{tabular}{|c|c|c|c|}
\hline
Class & Precision & Recall & F1-Score \\
\hline
Benign & 0.98 & 0.99 & 0.99 \\
Phishing & 0.96 & 0.90 & 0.93 \\
Defacement & 0.99 & 1.00 & 0.99 \\
Malware & 0.98 & 0.96 & 0.97 \\
\hline
Accuracy & & & 0.98 \\
\hline
\end{tabular}
\caption{Classification Report}
\end{table}

The model achieved an overall accuracy of 98\%, with the highest performance in detecting benign and defacement URLs. The relatively lower recall for phishing URLs suggests that further optimization is needed, particularly in the detection of phishing sites that closely resemble legitimate URLs.

\begin{figure}[h!]
    \centering
    \includegraphics[width=0.8\linewidth]{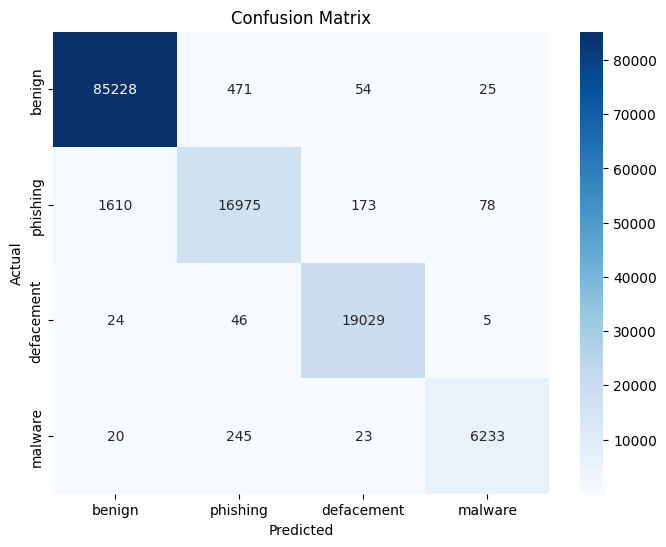}  
    \caption{Confusion Matrix of the Classification Model}
    \label{fig:confusion_matrix}
\end{figure}

\textbf{Rows} represent the actual class labels, and \textbf{columns} represent the predicted class labels. The matrix can be interpreted as follows:

- True Positives (TP): The number of correctly classified instances for each class.
- False Positives (FP): The number of incorrectly classified instances that were predicted as a given class but actually belong to another class.
- False Negatives (FN): The number of incorrectly classified instances that were predicted as not belonging to a given class but actually belong to that class.
- True Negatives (TN): The number of correctly classified instances that do not belong to a given class.

For each class, the confusion matrix provides:
\begin{itemize}
    \item Benign: 
    \begin{itemize}
        \item True Positives (TP) = 85,228 (Correctly classified as benign)
        \item False Positives (FP) = 471 (Incorrectly classified phishing URLs as benign)
        \item False Negatives (FN) = 54 (Incorrectly classified defacement URLs as benign)
        \item True Negatives (TN) = 25 (Correctly classified malware URLs as not benign)
    \end{itemize}
    \item Phishing:
    \begin{itemize}
        \item True Positives (TP) = 16,975 (Correctly classified as phishing)
        \item False Positives (FP) = 1,610 (Incorrectly classified benign URLs as phishing)
        \item False Negatives (FN) = 173 (Incorrectly classified defacement URLs as phishing)
        \item True Negatives (TN) = 78 (Correctly classified malware URLs as not phishing)
    \end{itemize}
    \item Defacement:
    \begin{itemize}
        \item True Positives (TP) = 19,029 (Correctly classified as defacement)
        \item False Positives (FP) = 46 (Incorrectly classified phishing URLs as defacement)
        \item False Negatives (FN) = 24 (Incorrectly classified benign URLs as defacement)
        \item True Negatives (TN) = 5 (Correctly classified malware URLs as not defacement)
    \end{itemize}
    \item Malware:
    \begin{itemize}
        \item True Positives (TP) = 6,233 (Correctly classified as malware)
        \item False Positives (FP) = 245 (Incorrectly classified phishing URLs as malware)
        \item False Negatives (FN) = 23 (Incorrectly classified defacement URLs as malware)
        \item True Negatives (TN) = 20 (Correctly classified benign URLs as not malware)
    \end{itemize}
\end{itemize}

This matrix helps identify the strengths and weaknesses of the model. For example, a high number of true positives (TP) indicates the model's effectiveness in correctly classifying instances. On the other hand, false positives (FP) and false negatives (FN) can highlight potential areas of improvement, such as misclassifying benign URLs as phishing or vice versa.

\section{Analysis}
There are existing models for phishing URL detection,
but this project introduces some new contributions that
can make the system more lightweight and accessible. The use of Bi-LSTM enables the model to learn both
forward and backward dependencies in URL structures,
resulting in a high detection accuracy of approximately
97\%, outperforming many baseline methods. The project also emphasizes real world deployment
through a lightweight Flask API and a user-friendly
interface. The interface also provides confidence scores for the
predictions which is a new contribution to the existing
models.

\section{Conclusion and Future Work}
In this paper, I proposed a deep learning-based approach for phishing URL detection using a Bi-LSTM model. This approach effectively classifies URLs into four categories: benign, phishing, defacement, and malware, and outperforms traditional methods in terms of accuracy and robustness.

Future work will focus on further improving the detection of phishing URLs by incorporating additional features, such as domain age, SSL certification, and user behavior.
The model also needs to work on optimizing the accuracy and train in a way that it can accurately predict unseen URLs. Additionally, the model can be extended to detect other forms of malicious URLs, such as those associated with spyware and ransomware.

\end{document}